\documentclass[10pt,iop]{emulateapj}
\usepackage{apjfonts}
\usepackage{epsfig}
\usepackage{amsmath}

\def\Meszaros{M\'esz\'aros~}

\begin{document}

\title{Modeling  the broadband emission of Fermi/LAT GRB 090902B}

\author{Ruo-Yu Liu\altaffilmark{1,2} and Xiang-Yu Wang\altaffilmark{1,2}}
\altaffiltext{1}{Department of Astronomy, Nanjing University,
Nanjing, 210093, China} \altaffiltext{2}{Key laboratory of Modern
Astronomy and Astrophysics (Nanjing University), Ministry of
Education, Nanjing 210093, China}


\begin{abstract}
GRB 090902B, detected by Fermi Large Array Telescope (Fermi/LAT),
shows extend high-energy emission ($>100$ MeV) up to $10^3$ s after
the burst, which decays with time in a power-law as $t^{-1.5}$. It
has been also observed by several follow-up low-energy instruments,
including an early optical detection  around 5000 s after the burst.
The optical emission at early time decays faster than $t^{-1.6}$,
which has been suspected to originate from the reverse shock. We
here explore the models that can possibly explain the the broadband
afterglow emission of GRB 090902B. We find that the reverse shock
model for the early optical emission would overpredict the radio
afterglow flux that is inconsistent with observations. A partially
radiative blast wave model, which  though is able to produce  a
sufficiently steep decay slope, can not explain the broadband data
of GRB 090902B. The two-component jet model, which consists of a
narrow and bright jet component in the core and a surrounding wider
and less energetic jet component, is shown to be able to explain the
broadband afterglow data, including the LAT high-energy data after
$\sim50$ s and low-energy (radio, optical and X-ray) afterglow data.
The early-time high-energy emission detected by LAT before $\sim50$
s is likely due to internal origin as that of the sub-MeV emission.
The highest energy (33 GeV) photon of GRB090902B detected at 80 s
can be marginally accommodated within the forward shock emission
under the optimistic condition that electrons are accelerated by the
Bohm diffusive shock.
\end{abstract}


\section{Introduction}
GRB 090902B, triggered the Fermi Gamma-ray Burst Monitor (GBM), is a
long, intense burst with a redshift of $z = 1.822$ (Cucchiara et al.
2009) and a fluence of $(4.36\pm0.06)\times 10^{-4} {\rm erg
cm^{-2}}$ (10 keV--10 GeV) over the duration $T_{90} \simeq 25$s of
the prompt emission. These data give an isotropic energy $E_{{\rm iso}} =
(3.63\pm0.05)\times10^{54} {\rm erg}$, comparable to the energy in
another bright GRB detected by Fermi, GRB 080916C (Abdo et al.
2009a). It was also detected by the Fermi LAT with extended high
energy ($>100$ MeV) emission up to 1000 s after the trigger, which
includes a 33.4 GeV photon at 82s after the trigger, the highest
energy yet detected from GRBs. The spectrum of this extended
emission is consistent with a power law with photon index
$\Gamma=-2.1\pm0.1$, and its flux ($>$100 MeV) declines as
$t^{-1.5\pm0.1}$ over the interval from 25 to 1000 s after the
trigger (Abdo et al. 2009b).

The optical-infrared afterglow of GRB 090902B has been observed by
several instruments (Pandey et al. 2010;  McBreen et al. 2010). The
earliest optical detection by ROTSE-IIIa occurring at about 5000 s
after the burst reveals a bright afterglow which decays faster than
$t^{-1.6}$ (Pandey et al. 2010). The temporal decay of the optical
afterglow becomes flatter after $\sim 12.5$ hours with a decaying
index of $-0.90\pm 0.08$. The X-ray afterglow, detected by Swift XRT
(Kennea \& Stratta 2009), is consistent with a single power law
decay with an index of $\sim 1.30\pm 0.04$ from 12.5 hours to 17
days after the burst. In the radio band, the light curve remains
flat until weeks after the burst and then starts to decline (Cenko
et al. 2010).

Kumar \& Barniol Duran (2010) modeled the the LAT, X-ray and
late-time ($t\ga 1{\rm days}$) optical data of GRB090902B with the
forward shock synchrotron emission and find that one forward shock
component can explain these data. However, their modeling does not
include the early optical data at 5000 s and the long-term radio
afterglow data. Cenko et al. (2010) model the late-time optical,
X-ray and radio data of GRB090902b with one forward shock component.
Following Pandey et al. (2010), they attribute the early optical
flash to the reverse shock emission. They also calculated the
expected high-energy emission using their best-fit parameters, but
find that the expected 100 MeV flux at $t=50$ s is a factor of 7
below the observed value. Given these incompleteness and
inconsistency, we aim at modeling the broadband data of GRB090902B,
including the LAT,  optical, X-ray and radio data.

One interesting feature of GRB090902B is the fast temporal decay of
the early optical emission, i.e. the power-law decay slope in
$F_\nu\propto t^{-\alpha}$ is $\alpha\gtrsim 1.6$. Analogous
phenomenon has been observed in GRB 990123 and has been interpreted
as arising from the reverse shock passing through the baryonic
ejecta of the GRB outflow, which produces an optical flash with a
steep decay (typically evolves as $t^{-2}$) at the early time
(M\'esz\'aros \& Rees 1997, 1999; Sari \& Piran 1999a, b; Panaitescu
\& Kumar 2004; Nakar \& Piran 2004, 2005). The reverse shock
scenario has been also proposed for the early optical emission of
GRB 090902B (Pandey et al. 2010; Cenko et al. 2010). With the
high-energy afterglow data from Fermi LAT for GRB 090902B, the
forward-reverse shock scenario for the all-band data can be further
tested.

There are other scenarios that can possibly explain the early fast
decay optical and high-energy emission, such as the fully or
partially radiative blast wave model (Ghisellini et al 20010; Wu et
al. 2005a). The radiation emitted by electrons accelerated in shocks
may lead to energy loss of the shock. The adiabatic shock
approximation will break down when the energy loss becomes
important, which occurs when the equipartition factor for electrons
(i.e. $\epsilon_{\rm e} $) and the radiation efficiency of electrons are
high. In the extreme case when almost all the shock energy goes into
electrons (i.e. $\epsilon_{\rm e} \simeq 1$) and the electrons are in the
fast-cooling regime, the shock becomes fully radiative. Although the
fully radiative shock is in general unrealistic for GRBs, a
partially radiative shock is possible when $\epsilon_{\rm e} $ is
relatively high, especially during the early afterglow phase. Due to
the decreasing energy remained in the forward shock, a faster decay
of the afterglow than that predicted by the standard scenario is
expected (Wu et al. 2005a). Motivated by this, we also consider
whether this scenario can explain the early fast decay of the
optical emission in GRB 090902B.

Numerical simulations show a structured outflow when the jet breaks
out the collapsing stellar envelope of massive stars (Zhang et al.
2003). In the structured jet model, it is usually assumed that the
energy per unit solid angle depends as a power-law or a Gaussian
function on the angular distance from the axis (e.g. \Meszaros,
Rees, \& Wijers 1998; Dai \& Gou 2001; Rossi, Lazzati, \& Rees 2002;
Zhang \& \Meszaros 2002; Kumar \& Granot 2003;  Granot \& Kumar
2003; Zhang et al. 2004). For the sake of  calculation ease, the
structured jet can be simplified as a two-component jet
(Ramirez-Ruiz et al. 2002, Berger et al. 2003; Peng et al. 2005;
Huang et al. 2004), which consists of a narrow and bright jet
component in the core and a surrounding wide and less energetic jet
component. When observers view along the axis, one will see the
afterglow emission produced by both the bright core and the broad
wings surrounding the core. As the narrow component has a small
opening angle, the jet break in the light curve can occur very early
and the afterglow emission produced by the narrow component will
have a fast decay after that time. This provides a potential
mechanism to produce the early-time fast decaying optical emission
of GRB 090902B, while  the late radio, optical and X-ray afterglows
that have normal light curves can arise from the wide component. We
will study whether this two-component scenario can explain the
broadband afterglow data of GRB 090902B.

We first study the reverse shock scenario in \S 2, and find that the
reverse shock scenario for the early optical emission predicts a
higher radio flux than the observed value.  In \S 3, we study
whether the partially radiative forward shock can explain the
broadband data of GRB 090902B, but find a negative result. Then we
propose a two-component jet model for the broadband afterglow data
of GRB 090902B in \S 4. In \S 5, we discuss the origin of early-time
high energy ($> 100$MeV) emission  observed during the prompt
bursting phase. Finally we give our conclusions and discussions in
\S 6.

\section{Forward shock - Reverse shock (FS-RS) model}
When the ultrarelativistic cold baryon-dominated GRB ejecta
encounters the cold ISM, a reverse shock that propagates back into
the ejecta and a forward shock that propagates into the ISM will
form, and as a result both the ejecta matter and the ISM matter are
heated up. An optical flash is expected to be  produced by the
reverse shock synchrotron emission, which should decay quickly due
to that the shocked ejecta is expanding adiabatically. The early
optical flash from GRB 990123 with a 9th magnitude is believed to be
such a good case (Sari \& Piran 1999a; \Meszaros \& Rees 1999). It
remains unclear why such optical flashes are lacked in general in
GRBs as ground-based robotic telescope observations (Yost et al.
2007) and early Swift UV/Optical Telescope (UVOT)  observations (Roming
et al. 2006) yield non-detections down to a much lower limit.

A bright optical flash is detected by ROTSE from GRB 090902B with a
magnitude of $m_R=16.4\pm0.5$ at  $t=5320$ s after the burst. The
optical non-detections at $2\times10^4$ s by ROTSE imply that the
optical emission decays faster than $t^{-1.6}$. Pandey et al. (2010)
and Cenko et al. (2010) argue that this rapid decay slope is
suggestive of the reverse shock origin. With the rich
multiwavelength observation data available for this burst, including
the high-energy LAT data, the radio  data, late-time optical and
X-ray data, we aim at testing this reverse-shock origin possibility.

\subsection{The forward shock emission}
The late-time optical and X-ray afterglow emission after half a day
shows a normal decay, with a decay slope of $\alpha_O=-0.90\pm 0.08$
and $\alpha_X\sim -1.30\pm 0.04$ respectively. They are broadly
consistent with the synchrotron afterglow emission produced by a
forward shock expanding into a constant-density medium with an
electron index of $p\simeq 2.2$ if the cooling frequency in the
synchrotron spectrum is between the optical and X-ray frequencies
(Kumar \& Barniol Duran 2010; Cenko et al. 2010). Following Sari et
al. (1998) and Wijers \& Galama (1999), we get the radius of forward
shock at time $T$
\begin{equation}
\begin{split}
R&=\left[\frac{17ET}{4\pi nm_{{\rm p}}c(1+z)}\right] ^{1/4}\\
&=4.05\times 10^{17}(1+z)^{-1/4}E_{54}^{1/4}n_{-3}^{-1/4}T_0^{1/4}{\rm cm},
\end{split}
\end{equation}
where $E$ is the isotropic-equivalent kinetic energy of the burst
and $n$ is the number density of the circum-burst ISM. Throughout
the paper, and we use c.g.s units and denote by $Q_x$ the value of
the quantity $Q$ in units of $10^x$.  Adopting $R=4\gamma^2cT$, the
bulk Lorentz factor of the shocked matter can be described by
\begin{equation}
\gamma =1838(1+z)^{3/8}E_{54}^{1/8}n_{-3}^{-1/8}T_0^{-3/8},
\end{equation}
Then we can obtain the characteristic frequencies of the afterglow
synchrotron emission
\begin{equation}
\nu_{{\rm mf}}
=1.05\times
10^{18}f^{2}(p)\epsilon_{e,-1}^{2}E_{54}^{1/2}\epsilon_{{\rm Bf},-5}^{1/2}(1+z)^{1/2}T_0^{-3/2}
{\rm Hz}
\end{equation}
where $f(p)=6(p-2)/(p-1)$, $\epsilon_{{\rm ef}}$ and $\epsilon_{{\rm Bf}}$ are
the equipartition factor for the energy in electrons and magnetic
field in the shock, and
\begin{equation}
\nu_{{\rm cf}}
=1.13\times 10^{24}E_{54}^{-1/2}n_{-3}^{-1}
\epsilon_{{\rm Bf},-5}^{-3/2}[1+Y(\nu_{{\rm cf}})]^{-2}(1+z)^{-1/2}T_0^{-1/2}
{\rm Hz}
\end{equation}
where $Y(\nu_{{\rm cf}})$ is Compton parameter for electrons that produce
the synchrotron photons at $\nu_{{\rm cf}}$. Hereafter the
superscripts/subscripts 'f' and 'r' are used to represent the
quantities of the forward shock and reverse shock respectively.  The
peak flux density is
\begin{equation}
F_{{\rm \nu,max}}^{\rm f}
=4.08\times
10^{-27}n_{-3}^{1/2}E_{54}\epsilon_{{\rm Bf},-5}^{1/2}(1+z)D_{{\rm L,28}}^{-2}
{\rm erg cm^{-2} s^{-1}}
\end{equation}
Then we get the flux density at a fixed frequency
\begin{equation}\label{Fnu}
F_{{\rm \nu}}^{\rm f}=F_{{\rm \nu,max}}^{\rm f}\left \{
\begin{array}{lll}
  (\nu/\nu_{{\rm mf}})^{1/3}  \propto T^{1/2} && \nu < \nu_{{\rm mf}} < \nu_{{\rm cf}},\\
  (\nu/\nu_{{\rm mf}})^{-(p-1)/2}  \propto T^{-3(p-1)/4} && \nu_{{\rm mf}} < \nu < \nu_{{\rm cf}},\\
  (\nu/\nu_{{\rm mf}})^{-p/2}(\nu_{{\rm cf}}/\nu_{{\rm mf}})^{1/2}\\ \propto T^{-(3p-2)/4}[1+Y(\nu)]^{-1} && \nu_{{\rm mf}} < \nu_{{\rm cf}} < \nu.
\end{array}
\right.
\end{equation}
where $Y(\nu)$ is the Compton parameter for the electrons whose
synchrotron frequency is $\nu$. Due to Klein-Nishina scattering
effect, $Y(\nu)$ is not a constant for the electrons that produce
high energy photons (Wang et al. 2010) and in fact is dependent of
$\nu$.

Now we confront the above theory with the observed data to constrain
the unknown parameters. From Abdo et al. (2009), Pandey et al.
(2010), we take the observed flux for LAT data at 173s, R-band
optical data at $1.30\times 10^5$s, X-ray data at $1.09\times 10^5$s
and radio data at $4.67\times 10^5$s as below:
\begin{equation}
\left \{
\begin{array}{llll}
F_{{\rm LAT}}^{\rm f}(173{\rm s})\simeq 3.74\times 10^{-31}{\rm erg cm^{-2} s^{-1}}\\
~~~~~~~~~~~~~~~(\nu_{\rm LAT} >\nu_{{\rm cf}}>\nu_{{\rm mf}}))\\
F_{{\rm X}}^{\rm f}(1.09\times 10^5{\rm s})\simeq 1.82\times 10^{-30}{\rm erg cm^{-2} s^{-1}}\\
~~~~~~~~~~~~~~~(\nu_{\rm X} >\nu_{{\rm cf}} >\nu_{{\rm mf}})\\
\kappa F_{{\rm opt}}^{\rm f}(1.30\times 10^5{\rm s})\simeq 9.33\times 10^{-29}{\rm erg cm^{-2} s^{-1}}\\
~~~~~~~~~~~~~~~(\nu_{{\rm cf}} >\nu_{\rm opt} >\nu_{{\rm mf}})\\
F_{{\rm radio}}^{\rm f}(4.67\times 10^5{\rm s})\simeq 6.83\times 10^{-28}{\rm erg cm^{-2} s^{-1}}\\
~~~~~~~~~~~~~~~(\nu_{{\rm mf}} >\nu_{\rm radio} >\nu_{af})\\
\end{array}
\right.
\end{equation}
where $\kappa \simeq 0.85$ is the extinction coefficient of the host
galaxy in R band (Cenko et al. 2010). The spectral regime of each
frequency is given in the parentheses.  Although $\nu_{{\rm LAT}}$ and
$\nu_{{\rm X}}$ belong to the same spectral regime, the X-ray flux and
high-energy flux data can give two independent constraints due to
two different $Y$ parameters. For electrons that produce the X-ray
synchrotron afterglow, the inverse-Compton scattering loss is in the
Thompson regime and $Y\sim
74.3\epsilon_{{\rm ef},-1}^{2/3}\epsilon_{{\rm Bf},-5}^{-4/9}E_{54}^{1/18}n_{-3}^{1/18}(1+z)^{1/18}T_0^{-1/18}$
in the slow-cooling phase, while for high-energy gamma-ray photons,
$Y$ parameter is small due to the KN suppression effect on the
scattering cross section, as we will show later.

So we obtain the constraints on the shock parameters
\begin{equation}
\left \{
\begin{array}{llll}
\epsilon_{{\rm ef},-1}^{6/5}\epsilon_{{\rm Bf},-5}^{1/20}E_{54}^{21/20}\simeq 12.9[1+Y(100 {\rm MeV})]\\
\epsilon_{{\rm ef},-1}^{6/5}\epsilon_{{\rm Bf},-5}^{4/5}E_{54}^{13/10}n_{-3}^{1/2}\simeq 51.6\\
\epsilon_{{\rm ef},-1}^{8/15}\epsilon_{1/2}E_{54}n_{-3}^{-1/18}\simeq 13.5\\
\epsilon_{{\rm ef},-1}^{-2/3}\epsilon_{{\rm Bf},-5}^{1/3}E_{54}^{5/6}n_{-3}^{1/2}\simeq 0.952\\
\end{array}
\right.
\end{equation}
where $Y(100 {\rm MeV})$ is the Compton parameter for the electrons
whose synchrotron frequency is $h\nu=100{\rm MeV}$ at $t=173$s. In
the calculation, a typical value of $p=2.2$ for forward shock has
been used. Solving the above equations, we obtain
\begin{equation}
\left \{
\begin{array}{llll}
E_{54}\simeq 2.0[1+Y(100 {\rm MeV})]^{3/4}\\
n_{-3}\simeq 0.40[1+Y(100 {\rm MeV})]^{1/4}\\
\epsilon_{{\rm ef},-1}\simeq 4.2[1+Y(100 {\rm MeV})]^{1/4}\\
\epsilon_{{\rm Bf},-5}\simeq 9.2[1+Y(100 {\rm MeV})]^{-7/4} .
\end{array}
\right.
\end{equation}
For the above set of parameters, we obtain $Y(100 {\rm MeV})\simeq
0.03f(p)^{-5/3}\epsilon_{{\rm ef},-1}^{-2/3}\epsilon_{{\rm Bf},-5}^{-1/3}E_{54}^{1/6}n_{-3}^{1/2}T_0^{1/2}(1+z)^{1/6}
\sim 0.06 $ at $t=173$s (Wang et al. 2010). As $Y(\rm 100 MeV)\ll
1$, we can just omit $Y(100 {\rm MeV})$ in these expressions.

\subsection{The reverse shock emission}
As the reverse shock propagates into the fireball ejecta, the number
of shocked electrons increases, which leads to an initial rise of
the reverse shock flux. Once it has passed through the shell, no new
electrons will be shocked  and the flux will decrease due to
adiabatic expansion of the radiating gas. The flux of reverse shock
peaks at $T_{\rm i}=\max(T_{{\rm dec}}, T_{{\rm cross}} )$ (hereafter, we use the
subscript 'i' to represent quantities at the peak time). Here
$T_{{\rm dec}}\equiv (l/\eta^{8/3}c)(1+z)$ is the shell deceleration time,
where ${l=(E/nm_p c^2)^{1/3}}$ is the Sedov length and $\eta$ is the
initial Lorentz factor of the ejecta.
$T_{{\rm cross}}\equiv(\Delta/c)(1+z)=T_{90}$, is the characteristic
timescale within which the reverse shock crosses the shell, where
$\Delta$ is the thickness of the shell (Sari \& Piran 1995, 1999b).
When the shell is thick so that $T_{{\rm cross}}>T_{{\rm dec}}$,  the reverse
shock becomes relativistic early on, and we call it  "thick shell"
case, otherwise it belongs to the "thin shell" case. Evolution of
bulk Lorentz factor, gas pressure and density of shocked electrons
after crossing time follows $\gamma \propto T^{-7/16}$, $p \propto
T^{-13/12}$ and $n \propto T^{-13/16}$ (Sari \& Piran 1999a, b;
Meszaros \& Rees 1999; Kobayashi 2000).

The bulk Lorentz factor of the  shocked shell at the peak time is
$\gamma_i =1838(1+z)^{3/8}E_{54}^{1/8}n_{-3}^{-1/8}T_{\rm i}^{-3/8}$. Then
we can get the Lorentz factor of the reverse shock at the peak time
\begin{equation}
\bar{\gamma}=\frac{1}{2}\left(\frac{\eta}{\gamma_i}+\frac{\gamma_i}{\eta}\right).
\end{equation} The
minimum random Lorentz factor of electrons is given by
\begin{equation}
\begin{split}
\gamma_{{\rm mr}}&=\epsilon_{{\rm er}}\frac{p-2}{p-1}\frac{m_p}{m_{\rm e} }(\bar{\gamma}-1)\\
&\simeq
16.7bf(p)\eta_{3}\epsilon_{{\rm er},-1}E_{54}^{-1/8}n_{-3}^{1/8}(1+z)^{-3/8}
T_{\rm i}^{3/8}\hat t^{-13/48}
\end{split}
\end{equation}
where $\epsilon_{{\rm er}}$ is the fraction of thermal electrons carried
by magnetic field in reverse shock,
$b\equiv\frac{(\bar{\gamma}-1)}{\eta /\gamma}$ (namely,
$\bar{\gamma}-1=b\frac{\eta}{\gamma}$) and $\hat t\equiv T/T_{\rm i}$.  We
use this approximation for $\bar{\gamma}-1$ so that we can treat the
calculation analytically. Usually, the value of $b$ is taken to be 1
(Waxman \& Draine 2000; Kobayashi et al. 2007) or $\frac{1}{2}$
(Wang et al. 2005). A mildly relativistic reverse shock
approximation with $\bar{\gamma}-1 \simeq 1$ is also sometimes used
(see e.g. Sari \& Piran 1999a). In this paper, we take
$b=\frac{1}{2}$ in the calculation, as $\eta\ga \gamma_i$. For a
typical value of $p=2.5$ for the reverse shock emission, $f(p)=2$.

We get the two characteristic frequencies for the reverse shock
synchrotron emission
\begin{equation}
\nu_{{\rm mr}}=9.24\times 10^{12}b^{2}f^{2}(p)\eta_3^2
\epsilon_{{\rm er},-1}^2\epsilon_{{\rm Br},-1}^{1/2}n_{-3}^{1/2}(1+z)^{-1}\hat
t^{-73/48} {\rm Hz}
\end{equation}
and
\begin{equation}
\nu_{{\rm cr}}=1.12\times
10^{18}E_{54}^{-1/2}n_{-3}^{-1}\epsilon_{{\rm Br},-1}^{-3/2}(1+Y_r)^{-2}(1+z)^{-1/2}T_{\rm i}^{-1/2}\hat
t^{1/16} {\rm Hz}
\end{equation}
The peak flux of the reverse shock synchrotron emission  is
\begin{equation}
\begin{split}
F_{{\rm \nu,max}}^{\rm r}=&9.75\times
10^{-22}\eta_3^{-1}E_{54}^{5/4}n_{-3}^{1/4}\epsilon_{{\rm Br},-1}^{1/2}\\
&(1+z)^{7/4}D_{{\rm L,28}}^{-2}T_{\rm i}^{-3/4}\hat
t^{-47/48} {\rm erg cm^{-2} s^{-1}} .
\end{split}
\end{equation}
Then we get the flux density at a certain frequency
\begin{equation}
F_{{\rm \nu}}^{\rm r}=F_{{\rm \nu,max}}^{\rm r}\left \{
\begin{array}{ll}
  (\nu/\nu_{{\rm mr}})^{1/3} ~~~~~ \propto T^{-17/36} & \nu < \nu_{{\rm mr}} < \nu_{{\rm cr}},\\
  (\nu/\nu_{{\rm mr}})^{-(p-1)/2}\propto T^{-(73p+21)/96} & \nu_{{\rm mf}} < \nu < \nu_{{\rm cr}}.\\
\end{array}
\right.
\end{equation}
If $\nu > \max(\nu_{{\rm mr}}, \nu_{{\rm cr}})$, the flux drops exponentially
with time because there are no new injected electrons whose typical
synchrotron frequency lies in this regime.

If we attribute the early optical emission at $t\simeq 5000$ s to
the reverse shock synchrotron emission, we would have
\begin{equation}
\kappa F_{{\rm opt}}^{\rm r}(5000{\rm s})= 0.91 {\rm mJy}
\end{equation}
where $\kappa \simeq 0.85$ is the extinction coefficient of the host
galaxy in R  band (Cenko et al. 2010). The reverse shock emission
can produce a radio flare as has been seen in GRB 990123 (Kulkarni
et al. 1999). Although the radio emission in GRB 090902B drops after
the first detection, it shows a flattening after the second
detection at $4.67\times 10^5{\rm s}$, so the forward shock emission
should be dominated at this time. Thus, we expect that
\begin{equation}
\begin{split}
F_{{\rm radio}}^{\rm r}(4.67\times 10^5{\rm s})&<
\frac{1}{2}F^{{\rm obs}}_{{\rm radio}}(4.67\times 10^5{\rm s})\\
&= 3.41\times
10^{-28}{\rm ergcm^{-2}s^{-1}}.
\end{split}
\end{equation}
At 5000 s, from equation (12) and (13) and using the constrained
value of $E$ and $n$ from the forward shock emission, we get
$\nu_{{\rm mr}}\sim 6.39\times
10^8\eta_3^2\epsilon_{{\rm er},-1}^2\epsilon_{{\rm Br},-1}^{1/2}$Hz and
$\nu_{{\rm cr}}\sim 3.29\times 10^{17}\epsilon_{{\rm Br},-1}^{-3/2}$Hz, so the
optical frequency $\nu_{{\rm opt}}$ ($4.7\times 10^{14}$Hz) is expected to
be in the regime $\nu_{{\rm mr}} < \nu_{{\rm opt}} < \nu_{{\rm cr}}$ and the flux
density should decline with a temporal index of $\simeq -2$. If the
radio frequency $\nu_{{\rm radio}}$ ($8.5\times 10^9$Hz) also lies in the
same regime (i.e. $\nu_{{\rm mr}} < \nu_{{\rm radio}} < \nu_{{\rm cr}}$) at
$4.67\times 10^5$s, the radio flux then relates with the optical
flux as
\begin{equation}
F_{{\rm radio}}^{\rm r}\simeq F_{{\rm opt}}^{\rm r} \left(\frac{4.67\times 10^5 \rm s}{5000
\rm s}\right)^{-2} \left( \frac{4.7\times 10^{14} \rm Hz}{8.5\times
10^9 \rm Hz}\right)^{-(p-1)/2}
\end{equation}
Combining Eqs.(17) and (18), we obtain $p<1.5$. Such a small $p$ is
inconsistent with the observed decay slope of the late-time optical
and X-ray emission. As  $\nu_{{\rm mr}}\propto \hat t^{-73/48}$, we note
that at $4.67\times 10^5$s, $\nu_{{\rm mr}}>\nu_{{\rm radio}}$ only if
$\eta_3>112\epsilon_{{\rm er},-1}^{-1}\epsilon_{{\rm Br},-1}^{-1/4}$. Since such
an initial Lorentz factor is too large, this spectral regime is
unlikely (see c.f. Ioka 2010).

Radio flux from the reverse shock may be also affected by the
synchrotron self absorption (SSA) of the radiating electrons. The
SSA frequency in the slow cooling case is (Wu et al. 2005a)
\begin{equation}
\nu_{\rm ar}=\left\{
\begin{array}{ll}
a^{3/5}\nu_{{\rm mr}},  \nu_{\rm ar} < \nu_{{\rm mr}},\\
a^{2/(p+4)}\nu_{{\rm mr}},  \nu_{{\rm mr}} <\nu_{\rm ar},
\end{array}
\right.
\end{equation}
where
\begin{equation}
a\equiv\frac{c_0(p-1)e\Sigma}{B_r\gamma_{\rm m} ^5}=377\eta_3^{-6}E_{54}n_{-3}^{-1/2}\epsilon_{{\rm er},-1}^{-5}\epsilon_{{\rm Br},-1}^{-1/2}(1+z)^2T_{\rm i}^{-2}t^{79/48},
\end{equation}
$c_0\simeq 15$ is nearly a constant and $\Sigma=\frac{E/\eta
m_pc^2}{4\pi R^2}$ is the column density  of electrons heated by the
reverse shock. {The SSA effect leads to a different light curve in
the radio band  from that in the optical band if $\nu_{{\rm radio}}$ lies
in one of the following regimes:
$\nu_{{\rm radio}}<\nu_{{\rm mr}}<\nu_{\rm ar}$,$\nu_{{\rm mr}} <\nu_{{\rm radio}}< \nu_{\rm ar}$,
and $\nu_{{\rm radio}}<\nu_{\rm ar}< \nu_{{\rm mr}}$. Among these three cases, the
second one is the most likely case for GRB 090902B since we expect
$\nu_{{\rm mr}}<\nu_{{\rm radio}}$ for normal parameter values. So we just
consider this case here and leave discussions on other cases in the
Appendix.} We find $\nu_{{\rm mr}}<\nu_{{\rm radio}}<\nu_{\rm ar}$ requires $
E_{54}n_{-3}\epsilon_{{\rm er},-1}\epsilon_{{\rm Br},-1}\ga 10^4$ at $4.67\times
10^5$s according to equations (3), (19) and (20). If we substitute
$E_{54}=2.0$ and $n_{-3}=0.40$, which are derived from the
constraints by the forward shock emission, into this inequality, we
obtain $\epsilon_{{\rm er},-1}\epsilon_{{\rm Br},-1}\gtrsim 1.25\times10^4$,
which is unlikely. Therefore we conclude that SSA effect can not
solve the problem that the reverse shock scenario for the optical
emission overpredicts the radio emission.

\section{Partially radiative model}
It is usually assumed that the blast wave that produces the
afterglow emission is nearly adiabatic (\Meszaros \& Rees 1997; Sari
et al. 1998), i.e. the total kinetic energy of the relativistic
shock is a constant. However, the energy loss of the blast wave
could be significant under some circumstance. The radiation
efficiency of the blast wave is given by (Wu et al. 2005a)
\begin{equation}
\epsilon =\left \{
\begin{array}{ll}
  \epsilon_{\rm e}  & \nu_{{\rm c}} < \nu_{{\rm m}},\\
  \epsilon_{\rm e}  \left(\frac{\nu_{\rm m} }{\nu_{\rm c} }\right)^{(p-2)/2} & \nu_{{\rm m}} < \nu_{{\rm c}}.\\
\end{array}
\right.
\end{equation}
As we can see from the above equation, the energy loss is especially
important in the "fast-cooling" case (i.e. $\nu_{\rm m}  >\nu_{\rm c} $) with a
large $\epsilon_{\rm e} $.  In this case, the decreasing blast wave energy
at early times will result in a faster decay of the afterglow
emission than the adiabatic case. At late time, as
$\frac{\nu_{\rm m} }{\nu_{\rm c} }$ decreases with time, the radiation efficiency
drops and the decay slope changes to the adiabatic case.  We examine
whether this scenario can explain the fast decay in the early
optical and high-energy LAT emission.

According to Huang et al. (1999) and Wu et al. (2005a), the
isotropic-equivalent energy $E$ of the blast wave evolve with time
as
\begin{equation}
E=E_0\left(\frac{T}{T_{{\rm dec}}}\right)^{-3\epsilon /(4-\epsilon)}.
\end{equation}
The quantities describing the synchrotron spectrum in such a
semi-radiative shock are similar to equations~(3)--(5), except that
the constant $E$ in these equations should be replaced by a
time-dependent $E$ as described by Eq.~(22). So the
synchrotron emission flux decays as $F_{{\rm \nu}}\propto
T^{-3(p-1+\epsilon)/(4-\epsilon)}$ for $\nu_{{\rm mf}}<\nu <\nu_{{\rm cf}}$. To
explain the decay slope $\alpha\ga 1.6$ of the early optical
emission in GRB 090902B with this model, we need $\epsilon \ga 0.6$.
Thus we need
\begin{equation}
\epsilon_{{\rm ef}}\ga0.6.
\end{equation}
To explain the late optical, X-ray  and radio observations, we also
need: $\kappa F_{{\rm opt}}^{\rm f}(10^5{\rm s})\simeq 0.01$mJy,
$F_{{\rm radio}}^{\rm f}(10^5{\rm s})\simeq 0.03$mJy, and $F_{{\rm X}}^{\rm f}(10^5{\rm
s})\simeq 0.2 \mu$Jy (radio observations starts at about $1.3\times
10^5$s, we extrapolate it to $10^5$s).  At $10^5$s, the two
characteristic frequencies in the synchrotron spectrum are
$\nu_{{\rm mf}}\simeq 2.0\times 10^{12} E_{54}^{1/2}\epsilon_{{\rm Bf},-5}^{1/2}
{\rm Hz}$ and $\nu_{{\rm cf}}\simeq 1.1\times
10^{17}E_{54}^{-11/18}n_{-3}^{-10/9}\epsilon_{{\rm Bf},-5}^{-11/18} {\rm
Hz}$ for $\epsilon_{\rm e} =0.6$ and $p=2.2$. As $\nu_{{\rm opt}}$, $\nu_{{\rm X}}$,
$\nu_{{\rm radio}}$ lie in three different frequency regimes, we have
three independent constraints on the shock parameters. Finally we
obtain
\begin{equation}
n_{-3}\simeq 0.53, \epsilon_{{\rm Bf},-5}\simeq 0.39, E_{54}(10^5s)\simeq
8.0
\end{equation}
For these parameters,  the deceleration time of the blast wave is
$T_{\rm dec}\simeq 120$s, and we can  obtain the blast wave kinetic
energy at the deceleration time $T_{\rm dec}$ according to
Eq.(\ref{ET}), i.e. $E_{54}(T_{\rm dec})\simeq 50$. This energy is
extraordinary large. With such a high isotropic energy, we expect
the flux density in LAT band at the deceleration time to be $F_{\rm
LAT}^{\rm f}(T_{\rm dec})\simeq 1\mu$Jy, which is one order of magnitude
higher than the observed flux ($\simeq 0.1\mu {\rm Jy}$). Therefore,
we conclude that this model can not explain the broadband data of
GRB 090902B.

\section{Two-component jet model}
Jets from GRBs may have complex structure. For the sake of
calculation ease, the structured jet can be simplified as a
two-component jet. It assumes that the jet consists of two
components: a narrow component with a relatively small half-opening
angle ($\theta_N$) and a large isotropic-equivalent energy  in the
center, and a wide component with a larger half-opening angle
($\theta_W$) and a smaller isotropic-equivalent energy (hereafter,
we use the superscripts/subscripts 'N' and 'W' represents the
quantities of the narrow component and the wide component
respectively). The energy distribution of the two-component jet can
be parameterized as:
\begin{equation}
E_{{\rm iso}}(\theta)=\left \{
\begin{array}{ll}
  E_{{\rm iso},N} & 0\leq \theta \leq \theta_N,\\
  E_{{\rm iso},W} & \theta_N \leq \theta \leq \theta_W\\
\end{array}
\right.
\end{equation}
The two components of the jet encounter the ISM and generate forward
shock and reverse shock respectively.  The observed afterglow
emission is the superposition of  the two component. Due to a higher
kinetic energy, the contribution by the narrow component to the
afterglow emission should dominate at the early time. As the bulk
Lorentz factor $\gamma_N$ of the narrow component decreases, the
light curves breaks to a steeper one once $\gamma_N < 1/\theta_N$.
After the break, the light curve of the narrow component declines
much faster than that of the wide one, and the contribution by the
wide component could dominate at late times. For the case of GRB
090902B, we can attribute the early optical emission around 5000s to
the forward shock emission of the narrow component after the jet
break, since the optical emission decays very fast. The extended
high-energy emission detected by LAT  can be attributed to the
afterglow emission of the narrow component before the jet break and
the late radio, optical and X-ray emission can be attributed to the
afterglow emission of the wide component.

Considering the inverse Compton loss of electrons,  the synchrotron
flux density at  frequency $\nu$ after the jet break is (Rhoads
1999; Sari, Piran \& Halpern 1999)
\begin{equation}
F_{{\rm \nu}}=F_{{\rm \nu,max}}\left \{
\begin{array}{lll}
  (\nu/\nu_{{\rm m}})^{1/3}  \propto T^{-1/3}\\
  ~~~~~~~~~~(\nu < \nu_{{\rm m}} < \nu_{{\rm c}}),\\
  (\nu/\nu_{{\rm m}})^{-(p-1)/2}  \propto T^{-p}\\
  ~~~~~~~~~~(\nu_{{\rm m}} < \nu < \nu_{{\rm c}})\\
  (\nu/\nu_{{\rm m}})^{-p/2}(\nu_{{\rm c}}/\nu_{{\rm m}})^{1/2} \propto T^{-p}[1+Y(\nu)]^{-1}\\
  ~~~~~~~~~~(\nu_{{\rm m}} < \nu_{{\rm c}} < \nu).
\end{array}
\right.
\end{equation}
The jet break of the narrow component should occur after the last
LAT detection (around 1000 s) and before the first ROTSE optical
data at about 5000 s. Assuming a jet break time at, e.g.
$T_{{\rm Br}eak}^{N}\sim 4000$s, the broadband data of GRB 090902B imply
\begin{equation}
\left \{
\begin{array}{lllll}
F_{{\rm LAT}}^N(173{\rm s})\simeq 3.74\times 10^{-31}{\rm erg cm^{-2} s^{-1}} \\
\kappa F_{{\rm opt}}^N(4000{\rm s})\left(\frac{5000 {\rm s}}{4000{\rm s}}\right)^{-p}\simeq 9.06\times 10^{-27}{\rm erg cm^{-2} s^{-1}}\\
\kappa F_{{\rm opt}}^W(1.3\times {\rm 10^5s})\simeq 9.33\times 10^{-29}{\rm erg cm^{-2} s^{-1}}\\
F_{{\rm X}}^W(1.09\times {\rm 10^5s})\simeq 1.82\times 10^{-30}{\rm erg cm^{-2} s^{-1}}\\
F_{{\rm radio}}^W(4.67\times {\rm 10^5s})\simeq 6.83\times 10^{-28}{\rm
erg cm^{-2} s^{-1}}
\end{array}
\right.
\end{equation}

By comparing the above flux data with the theory prediction of the
two-component jet model, we get the constraints
\begin{equation}
\left \{
\begin{array}{lllll}
\epsilon_{{\rm ef},-1}^{6/5}\epsilon_{{\rm Bf},-5}^{1/20}E_{\rm 54,N} ^{21/20} \simeq 12.9[1+Y(\rm 100 MeV)]\\
\epsilon_{{\rm ef},-1}^{6/5}\epsilon_{{\rm Bf},-5}^{4/5}E_{\rm 54,N} ^{13/10}n_{-3}^{1/2}\simeq 329\\
\epsilon_{{\rm ef},-1}^{6/5}\epsilon_{{\rm Bf},-5}^{4/5}E_{\rm 54,W} ^{13/10}n_{-3}^{1/2}\simeq 51.6\\
\epsilon_{{\rm ef},-1}^{8/15}\epsilon_{{\rm Bf},-5}^{1/2}E_{\rm 54,W} n_{-3}^{1/2}\simeq 13.5\\
\epsilon_{{\rm ef},-1}^{-2/3}\epsilon_{{\rm Bf},-5}^{1/3}E_{54,
W}^{5/6}n_{-3}^{1/2}\simeq 0.952 .
\end{array}
\right.
\end{equation}
Solving this set of equations, we obtain
\begin{equation}
\left \{
\begin{array}{lllll}
E_{\rm 54,N} \simeq 2.7[1+Y(\rm 100 MeV)]^{3/4}\\
E_{\rm 54,W} \simeq 0.65[1+Y(\rm 100 MeV)]^{3/4}\\
n_{-3}\simeq 0.28[1+Y(\rm 100 MeV)]^{1/4}\\
\epsilon_{{\rm ef},-1}\simeq 2.9[1+Y(\rm 100 MeV)]^{1/4}\\
\epsilon_{{\rm Bf},-5}\simeq 130[1+Y(\rm 100 MeV)]^{-7/4} .
\end{array}
\right.
\end{equation}
For the above parameter values, we find that $Y(\rm 100 MeV)\ll 1$
(Wang et al. 2010).

\begin{figure}[t]
\centering \epsfig{figure=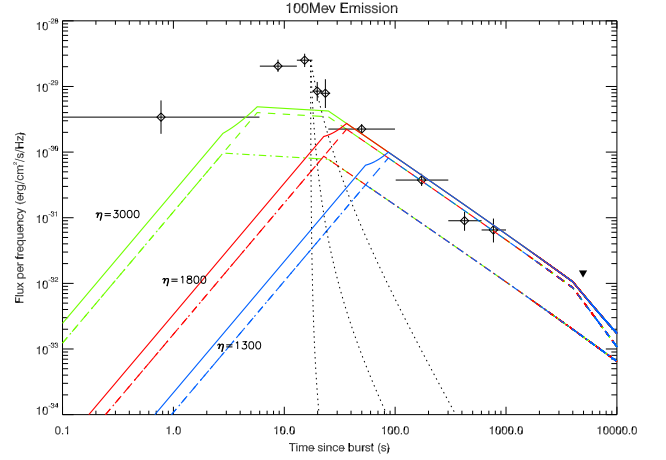,width=8.8cm} \caption{{ Fit of the
light curve of the high-energy ($>100 {\rm MeV}$) gamma-ray emission
from GRB 090902B with the two-component jet model. The dashed line
and dash-dotted line represent the forward shock synchrotron
emission of the narrow component and wide component respectively
(note that the two lines overlap during the rising phase), while the
solid line is sum of the contributions by two jet components. The
green, red and blue lines correspond to the initial Lorentz factor
of 3000, 1800 and 1300 respectively. The black dotted lines
represent the high-latitude emission of the last prompt LAT pulse,
calculated with Eq.(33), assuming pulse variability time of $\Delta
T=$0.05s, 1s, 5s from left to right respectively. Data are taken
from Figure 2 of Abdo et al. (2009b). The fitting parameters are
shown in Table 1.\label{fig1}}}
\end{figure}

\begin{figure}[t]
\centering \epsfig{figure=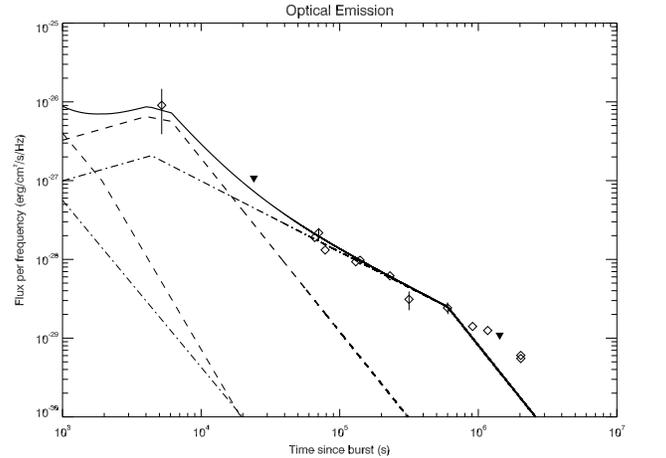,width=8.8cm} \caption{ Fit of the
light curve of the optical emission  from GRB 090902B with the
two-component jet model. The thick dashed line and dash-dotted line
represent the forward shock synchrotron emission of the narrow
component and  wide component respectively, while the solid line is
sum of the contribution by two jet components. The thin dashed line
and dash-dotted line represent the reverse shock  synchrotron
emission from the narrow component and wide component respectively.
Data are taken from Figure 6 of Cenko et al. (2010) and Figure 11 of
McBreen et al. (2010). The parameters used in the calculation of the
reverse shock emission is shown in Table 2. \label{fig2}}
\end{figure}

\begin{figure}[t]
\centering \epsfig{figure=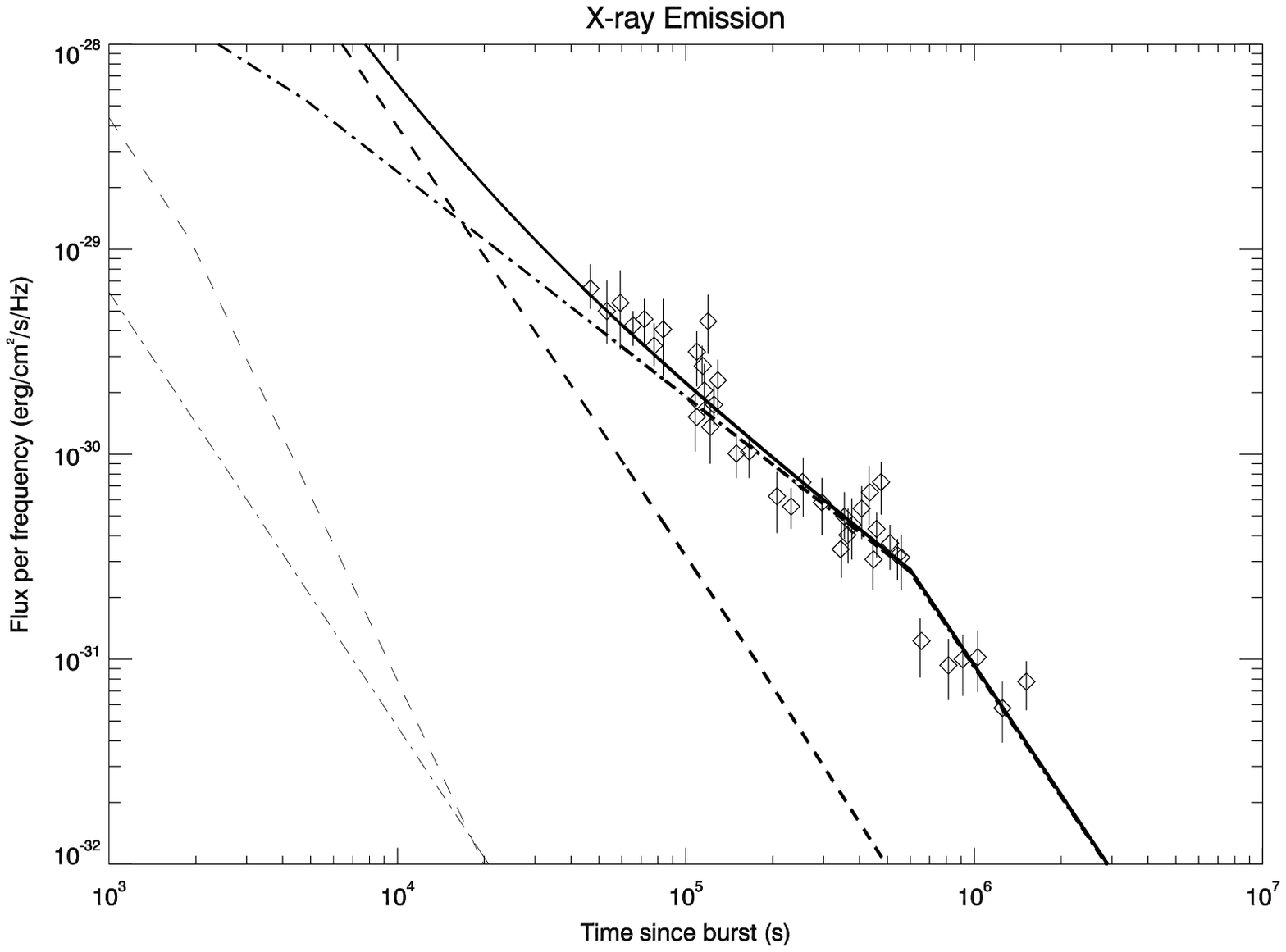,width=8.8cm} \caption{ Fit of
the light curve of the X-ray emission from GRB 090902B with the
two-component jet model.  All the lines represent the same as those
in Figure 2. Data are taken from Figure 6 of Cenko et al.
(2010).\label{fig3}}
\end{figure}

\begin{figure}[t]
\centering \epsfig{figure=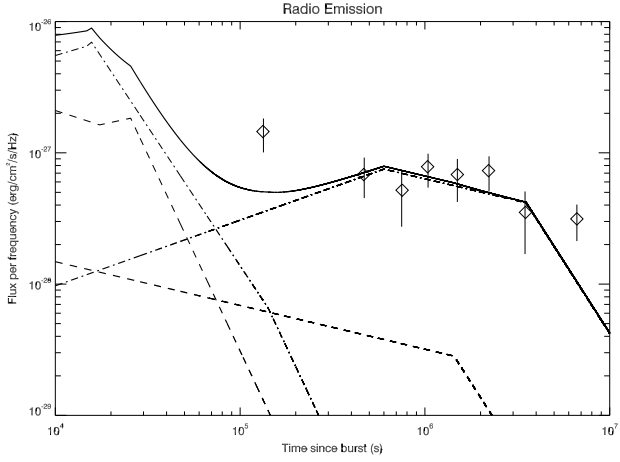,width=8.8cm} \caption{ Fit of
the light curve of the radio emission from GRB 090902B with the
two-component jet model. All the lines represent the same as those
in Figure 2. Data are taken from Figure 6 of Cenko et al.
(2010).\label{fig4}}
\end{figure}

Figures 1-4 show the fit result of the LAT, R band, X-ray and radio
data of GRB 090902B, respectively, and the fitting parameters are
given in Table 1. In the LAT energy band (Fig. 1), we plot the
forward shock emission for three different values of the initial
Lorentz factor, i.e. $\eta=3000$, 1800 and 1300 respectively. In the
case of $\eta=3000$, the shell crossing time $T_{\rm
cross}=T_{90}=25 {\rm s}$ is larger than the shell deceleration time
$T_{\rm dec}$, so the reverse shock belongs to the thick shell case.
In this case, the reverse shock will transit from the Newtonian
phase to the relativistic phase at time  ${t_{\rm N}=
\frac{{l^{3/2}}(1+z)}{{\Delta^{1/2}\eta^4 c}}=
0.38E_{54}^{1/2}n_{-3}^{-1/2}(\eta/3000)^{-4}(1+z)^{3/2}{\rm s}}$.
At time $t_{N}<t<T_{\rm cross}$, the Lorentz factor of the forward
shock evolves as $\Gamma\propto t^{-1/4}$, and the radius of the
shell evolves as $R\propto t^{1/2}$. Thus, the characteristic
frequencies and the peak flux evolve as $\nu_{{\rm mf}}\propto
t^{-1}$, $\nu_{{\rm cf}}\propto t^{-1}$ and $F_{\nu, max}^{\rm
f}\propto t$. As a result, the flux of high energy emission evolves
as $F_{\rm LAT}\propto t^{(-p+3)/2}$ in the regime $\nu_{{\rm
mf}}<\nu_{\rm LAT}<\nu_{{\rm cf}}$, and evolves as $F_{\rm
LAT}\propto t^{(-p+2)/2}$  in the regime $\nu_{\rm LAT}>\nu_{{\rm
cf}}$. For the inferred parameters of GRB090902B, $\nu_{\rm
LAT}>\nu_{{\rm cf}}$, so $F_{\rm LAT}\propto t^{(-p+2)/2}\propto
t^{-0.1}$ before $T_{\rm cross}$ (for $p=2.2$). We can see that the
$\eta=3000$  case (the green solid line) can not account for the
flux of the LAT peak at 10-20 s, although its light curve peaks at
the right time. When $\eta\la 1800$, the reverse shock belongs to
the thin shell case and light curve peaks after $T_{90}$. In the
thin shell case, the light curve rises initially as $t^2$ before the
peak. For $\eta=1800$, the forward shock emission can account for
the LAT data after $\sim 50$ s  (the red solid line), but falls
below the observed high-energy flux before $\sim 50$ s. When $\eta$
is smaller, the deceleration time is longer, as seen by the the case
of $\eta=1300$ (the blue lines).

In the optical band (Fig.2), the forward shock emission of the
narrow component reproduces the early optical flash at $\sim$ 5000
s. The isotropic energy in the narrow jet component is $E_{\rm N,
iso}=2.7\times10^{54}{\rm erg}$ and its half opening-angle is
$\theta_N=0.36^\circ$. The afterglow emission of the wide component
can also reproduce the late-time ($t\ga 1$ days) optical data as
well as the X-ray and radio data, if there is  a jet break at
$\sim6$ days (see Figs 2-4). The discrepancy between the model light
curve and the first radio detection at $\sim10^5$ s could be due to
strong interstellar scintillation of radio emission at early times
(Cenko et al. 2010). In the fitting, the half opening-angle of the
wide component is $\theta_N=2.8^\circ$ and the isotropic energy is
$E_{\rm W, iso}=0.65\times10^{54}{\rm erg}$. The circum-burst
density in the fitting is $n=0.28\times10^{-3} {\rm cm^{-3}}$, which
is at the low end of the ISM density{\footnote{Low density
circum-burst environment has also been found for other bright
Fermi/LAT GRBs such as GRB080916C, GRB090323 and GRB090328 (Cenko et
al. 2010). The reason why such bright Fermi/LAT GRBs have
preferentially low circum-burst density is  unknown. See Cenko et
al. (2010) for discussions on the possible explanations for such a
low density in these bursts.}}. Our fitting parameters of the wide
component agree well with that of Cenko et al. (2010), who modeled
only the late-time optical, X-ray and radio afterglow emission. With
the extra contribution by a narrow jet, our two-component jet model
is able to fit also the early optical data and the LAT high-energy
data after $\sim$ 50 s.  Due to a relatively low $\epsilon_{\rm e} $ and a
low circum-burst density $n$ that lead to a low radiative
efficiency, the radiative energy loss of the shock is unimportant,
so the adiabatic shock approximation used in the calculation is
valid.

Ryde et al. (2010) argue that a thermal photosphere component is
seen in the prompt burst emission of GRB 090902B. If this is true,
the composition of the jet should be baryon-dominated. In this case,
the reverse shock emission produced by the two components should
also be present. We calculate the reverse shock emission produced by
the two component jet, which are also  shown in Figures 1--4. We
find that the reverse shock emission is typically subdominant and
that the parameters of the reverse shock are not well constrained.
The parameter values given in Table 2 are just one illustration.

\section{Origin of the early-time high-energy ($>100{\rm MeV}$) emission}
We have shown that the forward shock emission of the narrow jet
component can account for the long-lived high-energy emission after
$\sim$ 50 s, as shown in Fig.1, but it can not account for the
early-time high-energy data.  The early LAT data is inconsistent
with a forward shock origin in the following aspects: 1)that the
high-energy emission of GRB090902B starts to decay before $T_{90}$
is unexpected because before $T_{90}$ the shell is putting energy
into the forward blast wave continuously to keep it from being
decelerated quickly; 2) the LAT peak at 10-20 s  is very sharp and
the decay slope immediately after the peak is too steep while the
forward shock emission model predicts a smooth and round peak; 3)
the temporal variability of the LAT emission during the prompt
bursting phase is correlated to some extent with the low-energy
emission detected by GBM, while the the forward shock emission model
predicts a smooth light curve. Therefore, there must be an extra
component that produces the high-energy emission before $\sim$ 50 s
in GRB 090902B. The same situation is seen in the modeling of
GRB090510 (He et al. 2010), in which the forward shock emission can
not account for the high-energy emission before $\sim 3$ s.

We first check whether the reverse shock can produce the peak at
10-20 s in the LAT light curve.   Due to that the flux of reverse
shock synchrotron emission rises rapidly with time before reaching
its peak and that the reverse shock operates only once, the light
curve of the reverse shock self inverse-Compton emission is expected
to form a peak at the crossing time (Wang et al. 2001a,b; Kobayashi
et al. 2007). However, due to the low density inferred for GRB
090902B from the broadband afterglow data, the scattering optical
depth of electrons in the ejecta  is low, so the IC flux is found to
be weak. For the parameter values given by Eq.(29), the scattering
optical depth in the reverse shock ejecta of the narrow component at
the crossing time is
\begin{equation}
\begin{split}
\tau_{r,N}&=\frac{\sigma_TN_{\rm e,N} ^{\rm r}}{4\pi R^2}=2.1\times 10^{-7}
\eta_{3}^{-1}E_{\rm 54,N} ^{1/2}n_{-3}^{1/2}(1+z)^{1/2}T_{\rm i}^{-1/2}\\
&\approx 6.3\times 10^{-8}\eta_3^{-1}
\end{split}
\end{equation}
where $N_{\rm e,N} ^{\rm r}=E_N/\eta m_pc^2$ is the number density of
electrons in the narrow component jet. According to Eq.(14), the
peak  flux of the reverse shock synchrotron emission is
$F_{max,N}^{r,syn}\approx 7.3\eta_3^{-1}\epsilon_{{\rm Br},-1}^{1/2}{\rm
Jy}$, then we get the peak flux of the SSC emission from the reverse
shock
\begin{equation}
F_{max,N}^{IC,rr}=\tau_r F_{max,N}^{r,syn}\sim
0.46\eta_3^{-2}\epsilon_{{\rm Br},-1}^{1/2}{\rm \mu Jy}.
\end{equation}
Since this peak flux is lower than the observed high-energy flux,
which is $\sim 3\mu {\rm Jy}$ (Abdo et al. 2009b), and the peak
frequency of the SSC emission locates at low energy, the SSC
emission from the reverse shock can not account for the LAT peak.
Using the parameters constrained by the low-energy broadband
afterglow data, we  find that the synchrotron flux of the reverse
shock at the LAT band is of the order of $10^{-2}\mu {\rm Jy}$,
which is also much lower than the observed value.

Therefore, the early LAT emission has to be attributed to some
internal dissipation process that occurs at radius much smaller than
the deceleration radius of the external shock.  We note that there
are pulses around 7.0s, 7.8s, 8.3s, 9.6s and 16s in LAT band, most
of which have corresponding GBM pulses. The coincidence supports the
viewpoint that the early-time LAT emission is due to internal
dissipation within the outflow, such as internal shocks. Ryde et al.
(2010) show that the time-resolved spectra of the prompt emission of
GRB090902B from KeV to GeV can be decomposed into two components,
one is the thermal multi-color blackbody component peaking at
sub-MeV and another is a power-law component extending into GeV.
Within this picture, the high-energy emission is attributed to the
non-thermal emission (synchrotron emission, SSC or Comptonization of
the thermal photons) produced by electrons in internal shocks (e.g.
Ryde et al. 2010; Toma et al. 2009; Pe'er et al. 2010). Hadronic
scenarios have also been proposed, including the proton synchrotron
radiation (Razzaque et al. 2010) or photohadronic cascade radiation
(Asano et al. 2009), which usually need a very large energy budget
due to the low radiation efficiency of protons (Wang et al. 2009).

The highest-energy (33 GeV) photon is detected after the prompt
burst, at 80 s after the trigger. In its local redshift ($z=1.822$)
frame, this energy is 94 GeV. Whether such a high-energy photon can
originate from the synchrotron radiation of the forward shock is an
interesting issue (Abdo et al. 2009b; Barniol Duran \& Kumar 2010;
Piran \& Nakar 2010), as the maximum energy of electrons is limited
by shock acceleration. At 80 s, the bulk Lorentz factor is
${\gamma\simeq 700}$ for the parameter values constrained by the
broadband afterglow data of GRB090902B. So the maximum synchrotron
photon energy is
\begin{equation}
\varepsilon_{\rm max}\simeq 40 \kappa^{-1}(\gamma/700){\rm GeV}
\end{equation}
according to Eq.(12) in Wang et al. (2009), where  $\kappa\ga1$ is a
parameter describing the efficiency of shock acceleration with
$\kappa=1$ corresponding to the fastest shock acceleration---the
Bohm diffusive shock acceleration in which the scattering mean free
path equals to the particle gyroradius. This means that the highest
energy photons of GRB090902B can be marginally accommodated by the
forward shock emission under the optimistic condition that particles
are accelerated by the Bohm diffusive shock. Another possibility is
that this highest energy photon belongs to the prompt component,
although the detailed radiation mechanism is unknown. In the latter
scenario, the deceleration time of the forward shock should be later
than 80 s, such as the $\eta=1300$ case (the blue lines) shown in
Figure 1.

Now we study the origin of the steep decay immediately after the LAT
light curve peak at $t=10-20$ s. A possible scenario for the steep
decay of the high energy photons is the high-latitude emission of
the jet at the end of the prompt emission phase. Because photons
from high latitude regions with respect to the line of sight will
arrive later than that from low latitude region due to the curved
front surface of the jet,  one observes a fast decreasing emission
from the high-latitude region rather than an abrupt stop of the
emission. The flux density of the high-latitude emission evolves
with time as (Kumar \& Panaitescu 2000; Wu et al. 2006; Toma et al.
2009)
\begin{equation}
F_{{\rm \nu}}(T)=F_{{\rm \nu}}(T_{\rm c} )\left(\frac{T+\Delta T-T_{{\rm c}}}{\Delta T}\right)^{-2-\beta}
\end{equation}
where $\Delta T$ is the duration of the last prompt pulse detected
by LAT, $T_{\rm c} $ is the LAT peak time  and $\beta$ is the spectral
index of high-energy photons. According to Figure 1 of Abdo et al.
(2009b), there is a LAT pulse around 16s. The variability time in
the energy band above 100 MeV is not well-determined, although the
sub-MeV emission detected by GBM shows variability timescale of
$\sim0.05$ s. In some internal dissipation models, such as the
residual internal shock model in which high-energy emissions arise
from much larger radii (Li 2010), the variability time in
high-energy emission could be longer than that of the low-energy
emission. Thus, we take $\beta=1.1$, $T_{\rm c} =16$s and assume three
different values for $\Delta T$ (i.e. $\Delta T=$0.05s, 1s and 5s
respectively) to test the high-latitude emission model. We calculate
the high-latitude emission according to Eq.(33), which is shown by
the black dotted lines  in Fig.1. As we can see, only with a very
large $\Delta T$ (i.e. $\Delta T\ga 5$s) can the peak-time data be
fitted by this model. As such a long variability is inconsistent
with the LAT data of GRB090902B, the steep decay may just reflect
the tail of the on-axis prompt emission. We note that a similar
conclusion has been reached for the early steeply decaying
high-energy emission in GRB090510 (He et al. 2010).

\section{Conclusions and Discussions}
With the broadband data from radio frequencies up to the LAT
high-energy ($\ga100{\rm MeV}$) band available, GRB 090902B is a
good case to examine the origin of the high-energy emission in GRBs
detected by Fermi/LAT. This burst has a bright optical flash
detected by ROTSE at $\sim$ 5000 s, which has been suspected to
arise from the reverse shock (Pandey et al. 2009; Cenko et al.
2010). We first try to fit the broadband afterglow observations of
GRB 090902B with the FS-RS  model, with the reverse shock emission
explaining the optical flash detected at 5000 s. We find that the
optical and radio observations cannot be explained simultaneously by
this model because a bright reverse shock optical emission will
yield a much higher radio flux than observed. The self-absorption
frequency of the reverse shock emission is found to be below the
radio band under the constraints  by the late-time forward shock
optical and X-ray emission, so the self-absorption effect can not
suppress the reverse shock radio emission.

Considering that the partially radiative blast wave scenario can
induce a fast decay of the afterglow emission,  we further test the
partially radiative blast wave scenario, but find that it can not
explain the broadband data of GRB 090902B either. This is mainly
because that late-time optical and X-ray flux constraints in
combination with the large energy loss in this scenario lead to a
huge initial kinetic energy in the blast wave which overpredicts the
early-time high-energy gamma-ray emission.

Then we propose that the two-component jet model, which consists of
a narrow and bright jet component in the core and a surrounding wide
and less-energetic jet component, is able to account for the
broadband observations of GRB 090902B. The long-lived high-energy
emission and early time optical emission can be attributed to the
forward shock synchrotron emissions of the narrow component. The
first optical detection at 5000 s should be later than the jet break
time of the narrow component so that a fast decay optical afterglow
is seen. From this, we derive that the half-opening angle of the
narrow component is $\theta_N\simeq0.36^\circ$. On the other hand,
the late-time optical, radio and X-ray afterglows can be attributed
to the wide component. To model the radio, X-ray and late-time
optical light curve, a jet break at $\sim$ 5-10 days is needed. From
this, we derive the initial half opening angle of the wide
component, $\theta_W\simeq 2.8^\circ$. The probability of observing
within the central bright core is only $\sim 10^{-2}$. This is
consistent with the rare detection of bursts with energy as large as
that of GRB090902B.

When observers view along the axis, the burst would be bright and
one will see the afterglow emission produced by both the bright core
and the broad wings surrounding the core. However, when observers
view the burst off-axis, one would miss the bright core and will see
the afterglow emission produced dominantly by the quasi-uniform
wing. As a consequence, we expect that the two-component jet
structure can be discerned more easily in brighter bursts, such as
GRB 090902B, through afterglow observations.  We note that the
two-component jet model is  favored  in another strong, long burst
-- GRB 080319B (Racusin et al. 2008). A two-component model has also
been invoked to explained the broadband data of a short GRB 090510
(Corsi et al. 2010), with  the wide component  explaining  a mild
excess in the optical band at late times \footnote{Using a
numerical code, He et al. (2010) find that a single jet model can
explain the broadband afterglow of GRB090510, except the early time
LAT data before $\sim 3$s.}. Extended high-energy emission is
especially useful to diagnose the narrow component, since at such
early time, high-energy emission is predominantly produced by the
narrow core component.

The early LAT emission of GRB090902B before $\sim$ 50 s after the
trigger can not be explained by the external shock emission and
should be due to an internal origin. This is consistent with the
multiple-pulse structure of the high-energy emission and its
temporal correlation with the sub-MeV emission seen during the
prompt bursting phase. Modeling of the broadband data of GRB090510
also show that the external shock emission cannot account for the
high-energy emission before $\sim 3$ s after the trigger (He et al.
2010).  The fact that one single Band function component can fit the
spectrum of the  prompt  emission from 10 keV to GeV in GRB080916C
also supports that the high-energy emission originate from the same
internal dissipation process as that of the sub-MeV emission (Abdo
et al. 2009a). Taken together, these results suggest that
high-energy emission of GRBs detected by Fermi/LAT during the prompt
bursting phase is dominated by the high-energy emission arising from
the internal origin, rather than the onset of the external shock,
although the external forward shock can readily account for the
extended high-energy after the prompt phase (Kumar \& Barniol Duran
2009, 2010; Ghirlanda et al. 2010).

\acknowledgments  We would like to thank Zhuo Li for valuable
comments and Zigao Dai, Yongfeng Huang and Haoning He for valuable
discussions. This work is supported by the NSFC under grants
10973008 and 11033002, the 973 program under grants 2009CB824800 the
Foundation for the Authors of National Excellent Doctoral
Dissertations of China, the Program for New Century Excellent
Talents  in University, the  Qing Lan Project and the Fok Ying Tung
Education Foundation.

\vskip 1cm
\appendix
{\center \bf Appendix: The forward-reverse shock scenario in
different spectral regimes}

As we mentioned in \S 2, there are three different regimes in
which the light curve in the radio band is affected by the
synchrotron self-absorption effect and  thus its behavior is
different from that in the optical band at $4.67 \times 10^5$s. In
fact, these three cases can be categorized into two cases, i.e. (1)
$\nu_{\rm ar}>\nu_{{\rm mr}}$ and $\nu_{\rm ar}>\nu_{{\rm radio}}$, (2)
$\nu_{\rm ar}<\nu_{{\rm mr}}$ and $\nu_{\rm ar}>\nu_{{\rm radio}}$.  We take
$E_{54}=2.0$ and $n_{-3}=0.40$, as constrained by the late radio, optical
and X-ray emission and a typical value of $p=2.5$  for the reverse
shock is used here.

\section{Thick shell case}
The necessary condition for the thick shell is $T_{90}>T_{{\rm dec}}$, so
we get $\eta_{3}\gtrsim 1.91$. According to equations (3), (19) and
(20), we have
\begin{equation}
\nu_{{\rm mr}}=1.05\times
10^6\eta_3^2\epsilon_{{\rm er},-1}^2\epsilon_{{\rm Br},-1}^{1/2}n_{-3}^{1/2}{\rm
Hz}
\end{equation}
and
\begin{equation}
a=5.13\times
10^7\eta_3^{-6}E_{54}n_{-3}^{-1/2}\epsilon_{{\rm er},-1}^{-5}\epsilon_{{\rm Br},-1}^{-1/2}
\end{equation}
at $4.67\times 10^5$s. So the SSA frequency is
\begin{equation}
\nu_{\rm ar}=\left \{
\begin{array}{ll}
4.52\times 10^{10}\eta_3^{-8/5}E_{54}^{3/5}n_{-3}^{1/5}\epsilon_{{\rm er},-1}^{-1}\epsilon_{{\rm Br},-1}^{1/5}{\rm Hz}, a<1\\
2.52\times 10^8\eta_3^{2/13}E_{54}^{4/13}n_{-3}^{9/26}\epsilon_{{\rm er},-1}^{6/13}\epsilon_{{\rm Br},-1}^{9/26}{\rm Hz}, a>1\\
\end{array}
\right.
\end{equation}

\subsection{$\nu_{\rm ar}>\nu_{{\rm mr}}$ and $\nu_{\rm ar}>\nu_{{\rm radio}}$}
According to the second equation of (A3), we get
$\eta_3^{2/13}\epsilon_{{\rm er},-1}^{6/13}\epsilon_{{\rm Br},-1}^{9/26}>37.4$
or $\eta_3>1.67\times
10^{10}\epsilon_{{\rm er},-1}^{-3}\epsilon_{{\rm Br},-1}^{-9/4}$ by requiring
$\nu_{\rm ar}>\nu_{{\rm radio}}$. Such a large value of the initial Lorentz
factor is unlikely, so we can exclude this case.

\subsection{$\nu_{\rm ar}<\nu_{{\rm mr}}$ and $\nu_{\rm ar}>\nu_{{\rm radio}}$}
By requiring $\nu_{\rm ar}<\nu_{{\rm mr}}$, we obtain
$\eta_3^6E_{54}^{-1}n_{-3}^{1/2}\epsilon_{{\rm er},-1}^5\epsilon_{{\rm Br},-1}^{1/2}>5.13\times
10^7$. On the other hand, from $\nu_{\rm ar}>\nu_{{\rm radio}}$ we obtain
$\eta_3^{8/5}E_{54}^{-3/5}n_{-3}^{-1/5}\epsilon_{{\rm er},-1}\epsilon_{{\rm Br},-1}^{1/5}<5.32$.
Combining these two inequalities together and taking $E_{54}=2.0$
and $n_{-3}=0.40$, we have $\eta_3<9.2\times
10^{-3}\epsilon_{{\rm Br},-1}^{3/4}$. Such a low initial Lorentz factor is
inconsistent with $\eta_3\gtrsim 1.91$, derived from the
pre-condition for the thick shell case,  so this case can be also
excluded.

\section{Thin Shell Case}
The thin shell case, on the contrary, requires $T_{{\rm dec}}\ga 25$s, or
$\eta_3\la 1.91$.  On the other hand, the fact that prompt
high-energy photons can escape from the source without annihilation
with low-energy photons puts a low limit on the initial bulk Lorentz
factor, which should be larger than a few hundreds. Another
constraint comes from the high-energy afterglow emission. If we
attribute the high-energy emission after 170 s detected by LAT to
the forward shock, the deceleration time should be shorter than 170s
and  then we get $\eta_3>0.92$. In the thin shell case, we have
\begin{equation}
\nu_{{\rm mr}}=6.43\times
10^6\eta_3^{-2}E_{54}^{1/2}\epsilon_{{\rm er},-1}^2\epsilon_{{\rm Br},-1}^{1/2}{\rm
Hz},
\end{equation}
\begin{equation}
a\approx 6.72\times
10^5\eta_3^{11/3}E_{54}^{-5/24}n_{-3}^{17/24}\epsilon_{{\rm er},-1}^{-5}\epsilon_{{\rm Br},-1}^{-1/2},
\end{equation}
and
\begin{equation}
\nu_{\rm ar}\approx \left\{
\begin{array}{ll}
2.02\times 10^{10}\eta_3^{1/5}E_{54}^{3/8}n_{-3}^{2/5}\epsilon_{{\rm er},-1}^{-1}\epsilon_{{\rm Br},-1}^{1/5}{\rm Hz}, a<1\\
3.99\times 10^8\eta_3^{-11/13}E_{54}^{6/13}n_{-3}^{3/13}\epsilon_{{\rm er},-1}^{6/13}\epsilon_{{\rm Br},-1}^{9/26}{\rm Hz}, a>1\\
\end{array}
\right.
\end{equation}
at $4.67\times 10^5$s respectively.

\subsection{$\nu_{\rm ar}<\nu_{{\rm mr}}, \nu_{\rm ar}>\nu_{{\rm radio}}$}
From $\nu_{\rm ar}<\nu_{{\rm mr}}$, we get $6.72\times
10^5\eta_3^{11/3}E_{54}^{-5/24}n_{-3}^{17/24}\epsilon_{{\rm er},-1}^{-5}\epsilon_{{\rm Br},-1}^{-1/2}<1$,
while from $\nu_{\rm ar}>\nu_{{\rm radio}}$, we get
$\eta_3^{1/5}E_{54}^{3/8}n_{-3}^{2/5}\epsilon_{{\rm er},-1}^{-1}\epsilon_{{\rm Br},-1}^{1/5}>0.42$.
Combining these two inequalities together, we have
$\eta_3^{8/3}<1.46\times 10^{-4}\epsilon_{{\rm Br},-1}^{3/2}$ or
$\eta_3<0.036\epsilon_{{\rm Br},-1}^{9/16}$. Since such a low initial bulk
Lorentz factor is inconsistent with the low limit on the initial
Lorentz factor constrained by the high-energy gamma-ray photons,
this case can be excluded.

\subsection{$\nu_{\rm ar}>\nu_{{\rm mr}}, \nu_{\rm ar}>\nu_{{\rm radio}}$}
Condition $\nu_{\rm ar}>\nu_{{\rm radio}}$ requires
$\eta_3^{-11/13}E_{54}^{6/13}n_{-3}^{3/13}\epsilon_{{\rm er},-1}^{6/13}\epsilon_{{\rm Br},-1}^{9/26}>21.3$.
If we substitute $E_{54}=2.0$ and $n_{-3}=0.40$ into this
inequality, we get
$\eta_3^{-11/13}\epsilon_{{\rm er},-1}^{6/13}\epsilon_{{\rm Br},-1}^{9/26}>19.1$
or $\eta_3<0.031\epsilon_{{\rm er},-1}^{6/11}\epsilon_{{\rm Br},-1}^{9/22}$.
Even with $\epsilon_{{\rm er},-1}=10$ and $\epsilon_{{\rm Br},-1}=10$, we still
have $\eta_3<0.28$. {Due to the same reason as the previous case,
this case can be ruled out either.}

Consequently, we conclude that either $\nu_{\rm ar}$ or $\nu_{{\rm mr}}$
cannot be above the radio  frequency  at the second radio detection
time $4.67\times 10^5$. The light curve in the radio  band at
$4.67\times 10^5$s should  behavior the same way as the optical
emission  around 5000s. And based on our discussion in \S 2, we
conclude that the FS-RS model is not viable for the broadband data
of GRB 090902B.

\clearpage

\begin{table}
\begin{center}
\caption{Forward-shock parameters in two-component jet model.\label{tbl-1}}
\begin{tabular}{cccccccccc}
\tableline\tableline
$E_{K,iso}^{narrow}$ & $E_{K,iso}^{wide}$ & $n$ &  $\epsilon_{{\rm ef}}$ & $\epsilon_{{\rm Bf}}$ & $\theta_N$ & $\theta_W$ & $p_f$  & $E_{jet}^{total}$\\
($10^{54}$ergs) & ($10^{54}$ergs) & ($10^{-3}$cm$^{-3}$) & ($10^{-1}$) & ($10^{-5}$) & ($\circ$) & ($\circ$) && ($10^{50}$ergs)\\
\tableline
2.7 & 0.65 & 0.28  & 2.9 & 130 & 0.36 & 2.8 & 2.2 & 8.3\\
\tableline
\end{tabular}
\end{center}
\end{table}

\begin{table}
\begin{center}
\caption{Reverse-shock parameters in two-component jet model.\label{tbl-2}}
\begin{tabular}{ccc}
\tableline\tableline
$\epsilon_{{\rm er}}$ & $\epsilon_{{\rm Br}}$ & $p_r$\\
($10^{-1}$) & ($10^{-1}$)\\
\tableline
3 & 1 & 2.5\\
\tableline
\end{tabular}
\end{center}
\end{table}

\end{document}